\documentclass[preprint,aps,preprintnumbers,amsmath,amssymb]{revtex4}
\usepackage{color}
\usepackage[]{graphicx}
\usepackage{latexsym}
\usepackage{natbib}
\usepackage{amsmath}
\usepackage[caption=false]{subfig}
\usepackage{multirow}
\usepackage{mathtools}
\usepackage{textcomp}
\usepackage{rotating}
\usepackage{tikz}
 
\usepackage{longtable}
\PassOptionsToPackage{hyphens}{url}
\usepackage[hyphens]{url}
\usepackage[breaklinks=true]{hyperref}

\usepackage[final]{feynmp}
\begin{document}

\title{Artificial intelligence for elections: the case of 2019
 Argentina primary and presidential election}

\author{Zhenkun Zhou}
	\affiliation{Levich Institute and Physics Department, City College of New York, New York, NY 10031, USA}
	\affiliation{State Key Lab of Software Development Environment, Beihang University, Beijing, 100191, China}

\author{Hern\'an A. Makse}
	\email{hmakse@ccny.cuny.edu}
	\affiliation{Levich Institute and Physics Department, City College of New York, New York, NY 10031, USA}

% \author[1,2]{Zhenkun Zhou}
% \author[2, *]{Hernán A. Makse}

% \affiliation[1]{State Key Lab of Software Development Environment, Beihang University, Beijing, 100191, China.}
% \affiliation[2]{Levich Institute and Physics Department, City College of New York, New York, NY 10031, USA.}
% \affiliation[*]{hmakse@ccny.cuny.edu}
\begin{abstract}

We use a method based on machine learning, big-data analytics, and
network theory to process millions of messages posted in Twitter to
predict election outcomes. The model has achieved accurate results in
the current Argentina primary presidential election on August 11, 2019
by predicting the large difference win of candidate Alberto Fern\'andez
over president Mauricio Macri; a result that none of the traditional
pollsters in that country was able to predict, and has led to a major
bond market collapse. We apply the model to the upcoming Argentina
presidential election on October 27, 2019 yielding the following
results: Fern\'andez 47.5\%, Macri 30.9\% and third party 21.6\%. Our method
improves over traditional polling methods which are based on direct
interactions with small number of individuals that are plagued by ever
declining response rates, currently falling in the low single
digits. They provide a reliable polling method that can be applied not
only to predict elections but to discover any trend in society, for
instance, what people think about climate change, politics or
education.

\end{abstract}

\maketitle

\section{Why pollsters are failing to predict elections}

Traditional polling methods \cite{surveys} using random digit dial
(RDD) phone interviews, opt-in samples of online surveys using, and
interactive voice response (IVR), are failing to predict election
outcomes across the world. For instance, the victory of Donald Trump
in the US 2016 presidential election came as a shock to many, as none
of the pollsters and political journalists, including those in Trump's
campaign, could predict this victory \cite{jacobs}.

%This fact made it clear that the traditional methods of surveys are
%not capturing the public opinions of the electorate. The failure to
%predict elections has been observed over and over in many cases: the
%most resonant perhaps was the election of Donald Trump in the US 2016
%Presidential election.

One of the reasons of this failure is that the percentage of response
to traditionally conducted surveys has decreased and it is becoming
increasingly difficult to get people's opinion \cite{aapor,pew}.
Response rates in telephone polls with live interviewers continue to
decline as has reached 6\% lower limit recently \cite{pew}. Response
rates could be even lower for other methodologies, like internet
polling or IVR. Thus, there is increasing evidence \cite{aapor,pew}
that the nonresponse bias might be the reason that polls are not
producing accurately matched election results.

The events leading to the recent primary election in Argentina on
August 11, 2019 are very telling about the failure of the polling
industry.

On the primary election day on August 11, 2019 (called PASO in
Spanish: Primarias, Abiertas, Simult\'aneas y Obligatorias; in
English: Open, Simultaneous, and Obligatory Primaries), none of the
pollsters in the country predicted the wide 16\% margin of triumph of
the Fern\'andez (formula Fern\'andez-Fern\'andez, FF) versus the
current president Macri (formula Macri-Pichetto, MP). The top five
pollsters Real Time Data, Poliarqu\'ia, Isonom\'ia, Giacobbe and
Elypsis showed the wrong result for the PASO election
\cite{erraron}. They all considered that Macri would win re-election
during their last minute predictions.
%The most active analyst was
%Luciano Cohan, economist and director of the consulting firm Elypsis
%\url{http://elypsisweb.com/},
%\url{https://ar.linkedin.com/in/luciano-cohan-a218a46}.
It was widely documented in the press (see Ref.~\cite{clarin}) 
that pollsters held several telephone conferences with foreign 
investors prior to the PASO election. This was the message: 
``Macri wins by one point: 38 to 37\%.'' 
It is important to consider that Macri was the right leaning
candidate favoring opening the economy to foreign investors and the
clear favorite of the market. Fern\'andez, on the contrary, was the
left-leaning candidate supporting cutting ties with foreign investors
and international markets.

As a result of the predictions supported by all Argentinian pollsters,
the bond market rose excessively in the days preceding the election
and generated circumstances in favor of Macri. With the lost of Macri
by 16 points, the bond market collapsed and the banks lost 1 billion
dollars at least. This speculative stumble caused another issue: on
the Monday after election day, the market overacted the fall and there
was a historic collapse of the Merval by 40\%. This catastrophic
collapse was coined the ``Lunes negro'' (Black Monday). Autonomy
Capital's founder, Robert Gibbins, known for making concentrated
investments in foreign markets, was one of the expert hedge funders
who lost the most in this market collapse. As documented in the Wall
Street Journal: {\it ``Hedge Fund Loses \$1 Billion in One Month on
 Argentina Bet''} \cite{wsj}, 
Autonomy Capital lost about \$1 billion largely on investments tied to
the Argentina collapse, making it one of the most prominent investors
caught on the wrong side of market turmoil in that country. Bankers
defended themselves for being on the wrong side of the story by saying
that they would not have bet so much if the Casa Rosada (the Macri
administration) had not endorsed the unreal data of the
pollsters.

Most of these polls employ a combination of IVR of landline numbers
complemented with online opt-in samples from Facebook. The vast
majority of these online panels in Argentina as well in US are made up
of volunteers who were recruited online and who receive some form of
compensation for completing surveys, such as small amounts of money or
frequent flyer miles.

In this paper, we present a method based on AI and big-data from
Twitter Search API \cite{bovet} that correctly predicted the results
of the Argentinian primary election on August 11, 2019. We also
present our predictions for the upcoming general presidential election
in that country which are: Fern\'andez 47.5\%, Macri 30.9\% and 
third party 21.6\%.

\section{Artificial Intelligence for Elections}

In this paper, we present a method based on Artificial Intelligence
(AI), which tries to identify political and electoral trends without
directly asking people what they think, but trying to predict and
interpret the enormous amount of data that people produce in the
digital age \cite{bovet,fake,book,jasny}. In fact, in recent years, the
digital revolution has allowed citizens of the world to express their
opinion openly on platforms such as Twitter and others. We will use
this information to understand the trends in public opinions of
society and to monitor the opinion of the electorate regarding
candidates in any particular election. With current digital platforms
that house information from millions to billion of people, compared to
traditional surveys that do not exceed a thousand, we pose that the
future of electoral forecasts is in the use of AI and Big-Data.

Big Data, in fact, is processing real data in real time to discover
where public opinion is heading in political, social and economic
matters. However, it is well known that social networks generate a
huge amount of spurious, false, erroneous data through trolls, bots
and misinformation campaigns. By virtue of this, the great challenge
of algorithms and AI is to discover and interpret real data from 'junk
data' that could lead to accurate predictions of electoral or opinion
trends. This feat is realized by the use of machine learning.

Below, we present a AI model, that has been already tested in the US
elections \cite{bovet,fake} and, recently in Argentina, to extract the
real opinion of people on social networks via algorithms that use
machine learning. We compare results to the traditional polls during
the Argentina election. The results show that AI can capture the
public opinion more precisely and more efficiently than traditional
polls.

\section{Methods}

We use machine learning and large-scale data (ie big data), from
social networks such as Twitter, to deduce the opinion of millions of
users who discuss politics and share their opinions via social
networks.

The first step of the method is to collect a large number of tweets
and make a basic statistical analysis of what users are talking
about. We use the following queries based on the names of the
candidates to the 2019 Argentina primary election: 
{\it Alberto AND Fernandez}, 
{\it alferdez},
{\it CFK},
{\it CFKArgentina}, 
{\it Kirchner},
{\it mauriciomacri},
{\it Macri}, 
{\it Pichetto},
{\it MiguelPichetto},
{\it Lavagna}.
Using these queries we have collected 45 million tweets
from 2 million users regarding the elections in Argentina from March
2019 until the day of publication, October 24, 2019. This large amount
of tweets collected has no precedent and is relevant in the light of
considering that Argentina is one of the most tweeting countries in
the world.

\begin{figure}[h!]
 \includegraphics[width=\textwidth]{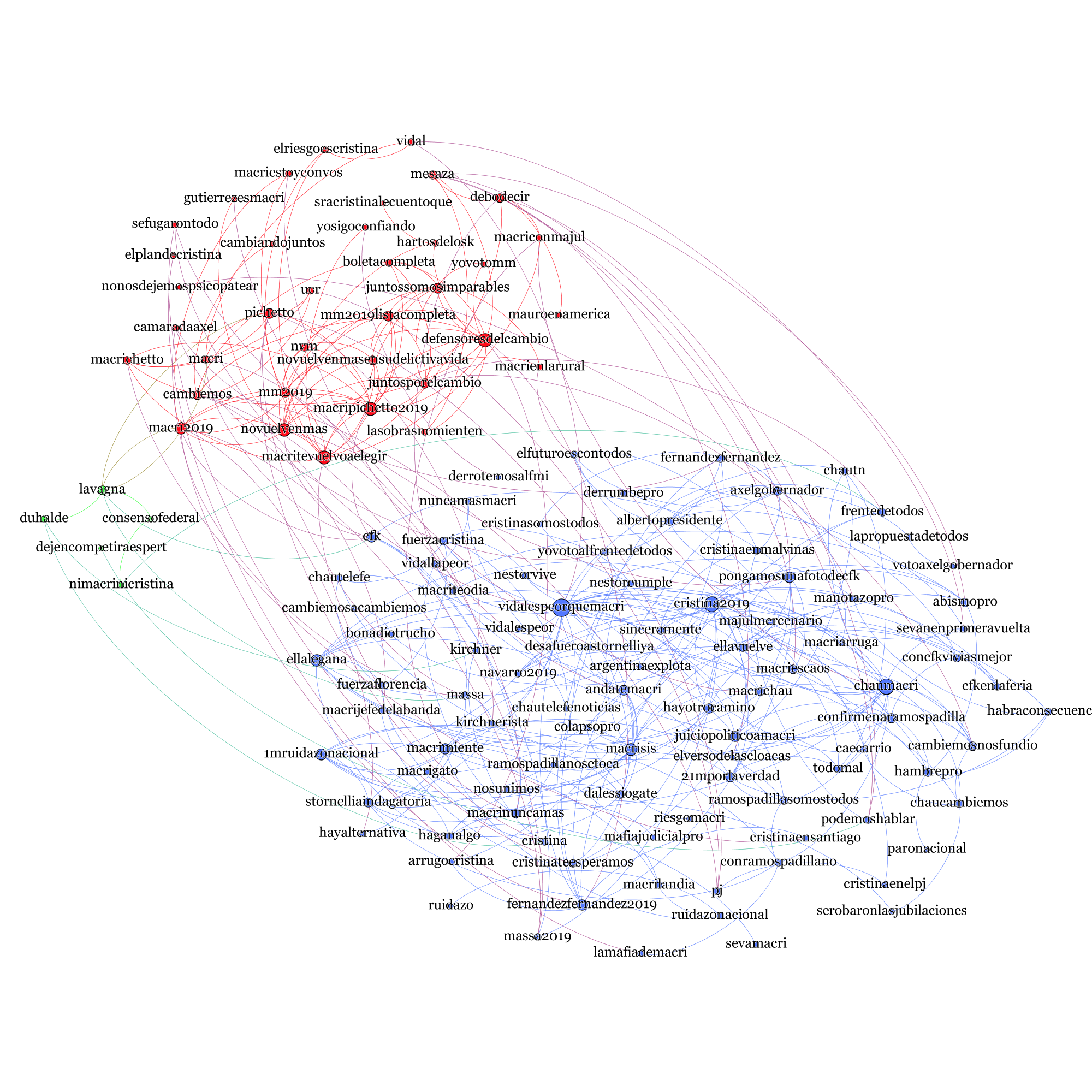}
\caption{Hashtag co-occurrence network from March to August 2019. Three groups of hashtags defined among the users: blue hashtags in favor of Cristina Fernandez de Kirchner, red hashtags in favor of Macri and a very small group in favor of the third candidate Lavagna colored in green.}
\label{hashtags}
\end{figure}

After discarding all bots \cite{bovet}, we proceed to analyze the
content of the tweets and identify relevant hashtags. A first simple
graph analysis is to order the most important hashtags to understand
the opinion of these users as shown in Fig. \ref{hashtags}.

We can see how there are three groups defined among the users: those
in favor of Cristina Fernandez de Kirchner (the vice-president
candidate), those in favor of Macri (the incumbent) and a very small
group in favor of the third candidate Lavagna. This analysis gives a
first electoral panorama that indicates that it is clear that the
followers in favor of Cristina Kirchner are much more passionate than
those of Macri. For example, Kirchner's type of hashtags is
\#FuerzaCristina \#Nestorvuelva \#Nestorpudo or they are very negative
to Macri as \#NuncamasMacri. On the other hand, Macri's group is much
smaller and less passionate with hashtags like \#Cambiemos or
\#MM2019, while support for the third candidate has not taken traction
and its electoral base on Twitter is very small.

This first analysis gives Cristina Kirchner a large advantage. In
fact, making a simple account of the number of users of the hashtags
would give a prediction that Fernandez-Fernandez would win. But a
correct prediction has to consider all users, not just the 10 thousand
users who express their opinion through hashtags.

AI then allows a machine to do in seconds what a human would take
hundreds of years. For example, reading and classifying into each
party camp each of the collected 45 million tweets, the AI can do in a
few minutes. For this purpose we train a machine learning model that
reads every tweet that users write and then predict the meaning of the
tweet and classify the tweet as favorable to the each one of the
candidates. Machine learning allows us to classify in seconds who is
the follower of Macri, Fernandez or Lavagna.

Following this logic, we have previously achieved surprising results
from the presidential elections in the United States in 2016~\cite{bovet}. 
We next apply the methods to understand the dynamics of the presidential elections in Argentina.

\section{Instantaneous and Cumulative Predictions in PASO elections in Argentina}

\begin{figure}[h!]
 \includegraphics[width=.9\textwidth]{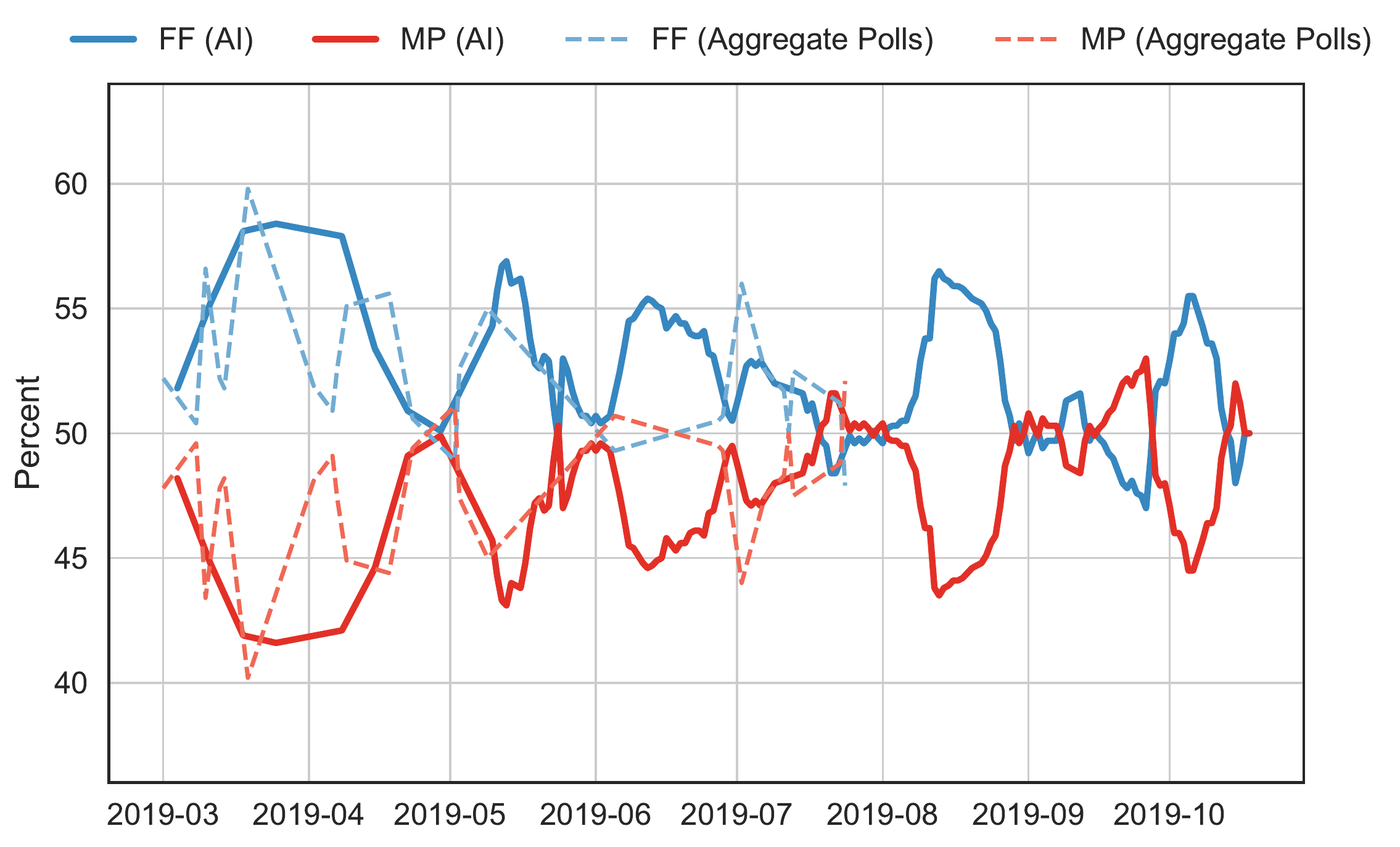}
 \caption {Instantaneous prediction ($w=14$) and trusted polls for two candidates.}
\label{fig1}
\end{figure}

\begin{figure}[h!]
 \includegraphics[width=.9\textwidth]{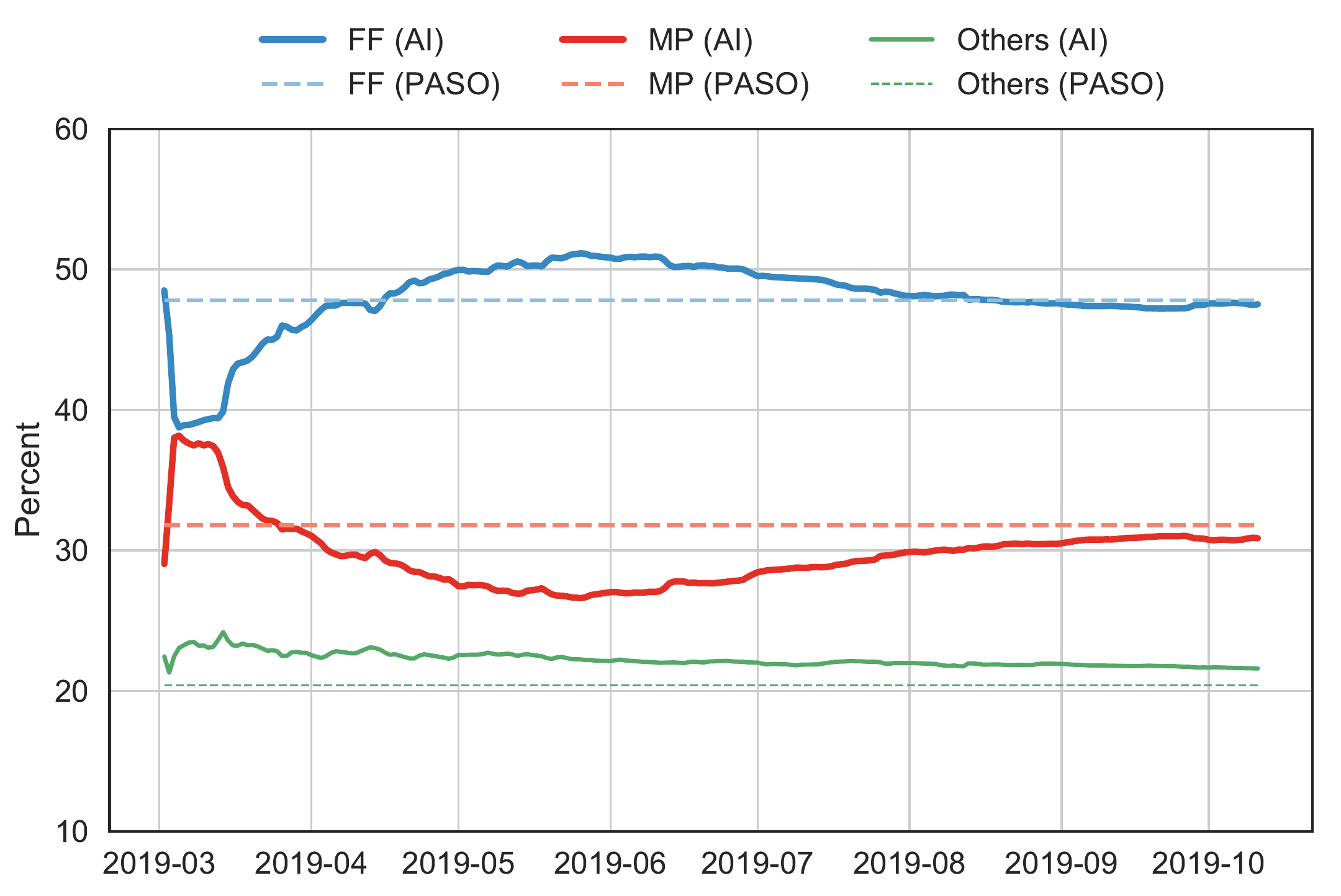}
 \caption {Cumulative prediction from the time of initial measurement in March 2019. The cumulative indicator accurately captures the results of the primary elections (PASO) on August 11, 2019.}
\label{fig2}
\end{figure}

We develop two indicators for opinion trends, see Figs \ref{fig1} and
\ref{fig2}. In the first we obtained a snapshot of the opinion in a
time window of $w$ days. In our studies we set $w=14$ days, but this
parameter can be changed as needed. The results of this instantaneous
predictions for the Argentinian election can be see in
Fig. \ref{fig1}.

This instantaneous indicator has been used in our previous study of
the 2016 US presidential election providing a very accurate fitting to
the results of the New York Times Aggregator of Polls at `The
Upshot' \cite{nytimes}. This aggregator assemble a large number of polls, of the order
of thousands, and weight them with proprietary information to produce
a weighted average of all the most trustable pollster in USA. In
Fig. \ref{fig1} we also plot the results of the most trusted polls in
Argentina as published by newspapers like Clar\'in and La Naci\'on and
also compiled at Wikipedia for a comparison with our instantaneous
prediction \cite{wiki_election}. Similarly to the case of the NYT and The Upshot
predictions, our AI follows the traditional polls in an average way.

For instance, using this snapshot, in April 2019, our model had
predicted the collapse of the image of Macri and the increase of
Cristina Kirchner as a possible candidate for the presidency, a
prediction that have agreed with an important pollster in Argentina,
Poliarquia, whose polls produced a large movement in financial
markets.

While these analyzes are interesting and give the opportunity to
predict instantaneous changes in electoral opinion, this indicator
does not provide the electorate opinion as a whole and it is not the
most important predictor of the election outcome. It is not the
greatest information that can be extracted from social networks,
either. Such a predictor is provided when we consider the cumulative
number of users from the beginning of measurements, and not just the
behavior of the users in a small window of information.

The fundamental difference between the analysis of social networks and
traditional surveys is that one can track millions of new users in the
networks and follow them over much longer periods that can cover
months or the entire election season. This cannot be done with
traditional surveys because the representativeness of the respondents
is less (less than a thousand people), in addition to the fact that
the statistical ensemble of the people who are surveyed changes
completely every week with each survey.

%In social networks, on the other hand, people openly express their
%opinion about the country's problems, including the economy, security,
%corruption and all kinds of issues that afflict them. To obviate
%people giving their opinion on social networks is a waste for anyone
%who wants to understand the reality in which we live. In fact, the
%great face-to-face manifestations of the past are diminishing to give
%way to social media protests. 

Our model can monitor not only the population's response to daily
events in a short window of observation $w$, but also the cumulative
opinion of each user for a prolonged period of months or years, making
trends in favor or against a candidate evident, see SM Section
\ref{def1}. Figure \ref{fig2} shows the results of the cumulative
opinion from the time of initial measurement in March 2019. We find
that this cumulative indicator captures the results of the 
primary elections on August 11, 2019
(PASO, in Spanish: Primarias, Abiertas, Simult\'aneas y Obligatorias; 
in English: Open, Simultaneous, and Obligatory Primaries).

This crucial point is evident when we compare the prediction of AI
considering the weekly opinion of users on Twitter in Fig. \ref{fig1}
with the opinion accumulated since March obtained in
Fig. \ref{fig2}. For example, as seen in Fig. \ref{fig1} in the last
month before the elections, the Macri-Pichetto formula had approached
the Fernández-Fernández predictions but, as of August 1 and throughout
the last week of the elections, the seconds took a distance
considerable and reached 7 percentage points above Macri-Pichetto. So
far, this weekly prediction does not explain the great difference of
16 points obtained in the primary elections.

Only by capturing user opinion data and normalizing the data to
include the representative demographics of society is that one obtains
the remaining 8\%. In this scenario, it is clear that AI allows to
capture data from social networks more accurately and for an extended
period of time (much more than one or two weeks). Figure \ref{fig2}
clearly demonstrates that considering the opinions of all Twitter
users during the five months of the campaign, the large difference
is captured and then reflected in the elections.

It is only until we analyze the trend of cumulative results during the
last five months that it becomes clear that the Fernández-Fernández
formula had always been above Macri-Pichetto. This cumulative
prediction is the best predictor of the election and accurately captures the difference that was observed at the primary election. 
On the other hand, the instantaneous snapshot over 
a finite window of time $w$
remains a better indicator for the opinion polls.

\section{Opinion of the electorate}

\begin{figure}[h!]
  \includegraphics[width=.8\textwidth]{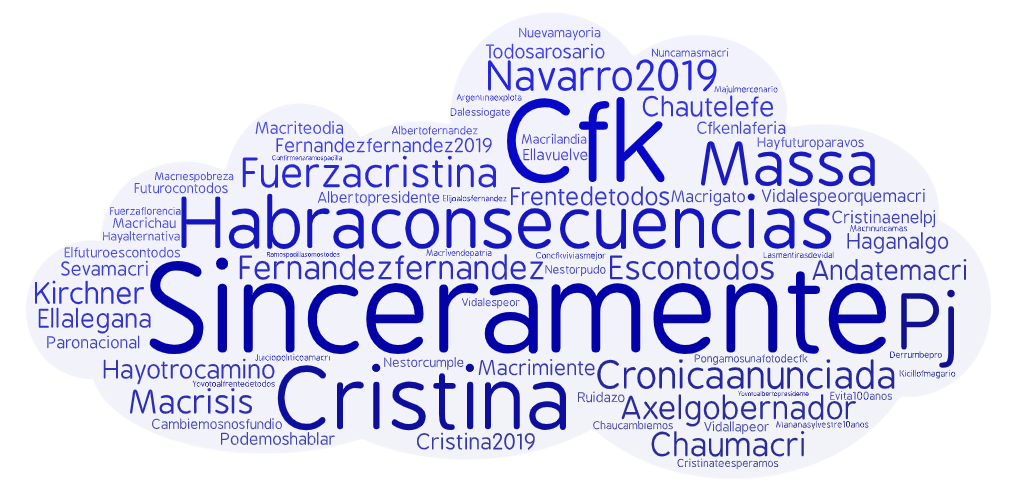}
  \includegraphics[width=.8\textwidth]{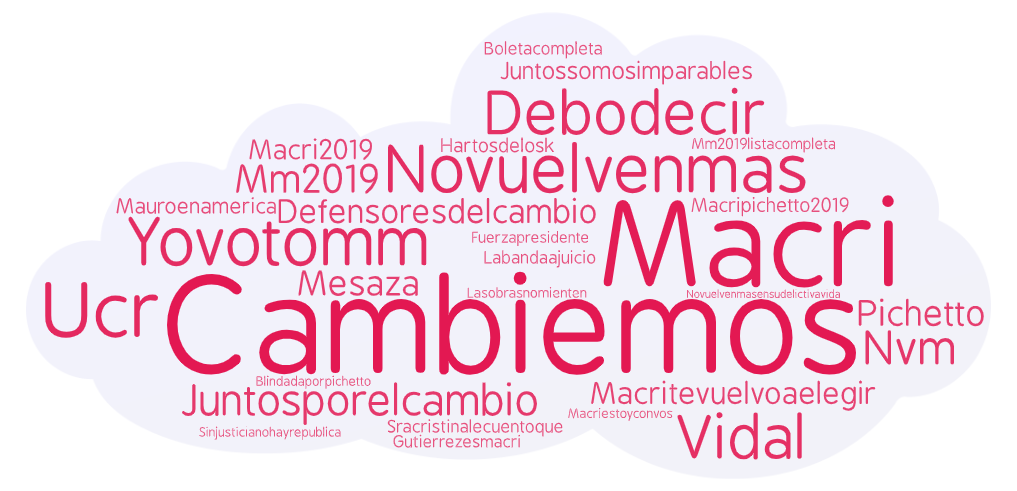}
 \caption{Hashtag clouds. The blue hashtags are most frequently used in the tweets in favor of Cristina and the red hashtags are those in favor of Macri.}
 \label{opinion_trends}
 \end{figure}

The AI allowed to capture the feelings of public opinion as seen in
Fig.~\ref{opinion_trends}. In the primary elections, a study of the hashtags and queries of the followers of the Fernández-Fernández formula indicates
that the vast majority did not stop thinking of the judicial cases of
corruption that affect Kirchenerism, but instead focused on expressing
anger at the poor economic situation in which the country is. In most
hashtags, there is great anger that reflects the hunger, chaos, crisis
and despair that afflicts society. On the other hand, the expression
of the followers of Macri-Pichetto is reflected in hashtags to give
strength to the president but they do not reflect a feeling for the
economic and political situation, but more a moral support, perhaps
of resignation. The followers of Macri do not express too much their
concerns about judicial cases of corruption. The third position never
managed to crystallize in the Twitter sphere.

\section{Predictions for the General Elections on October 27, 2019}

\begin{figure}[h!]
 \includegraphics[width=\textwidth]{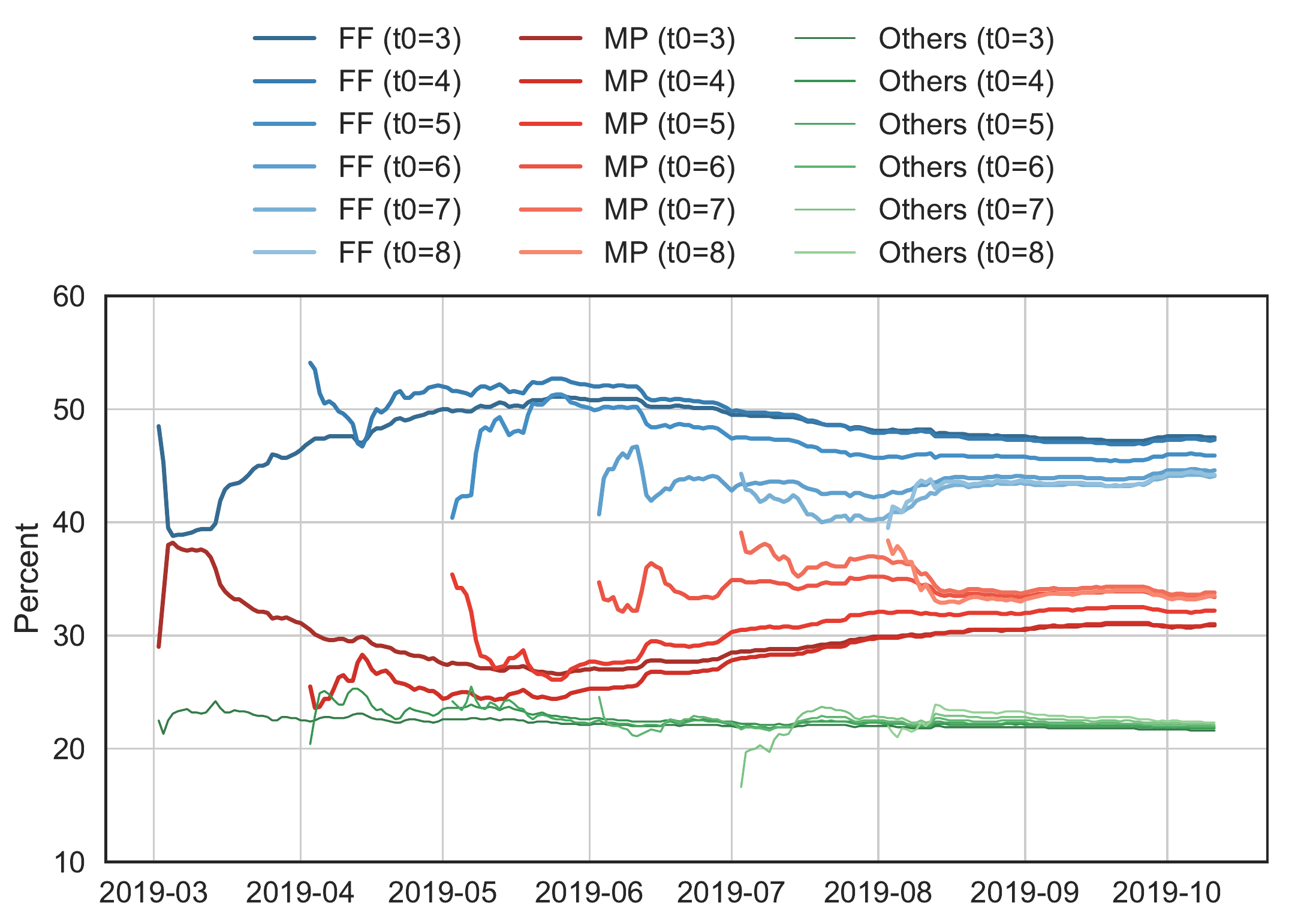}
\caption {Cumulative prediction for different initial times $t_0$ (3 for March, 4 for April, etc.).  While we consider different $t_0$, the predictions at the day of the election clusters well around the results of the PASO.}
\label{fig3}
\end{figure}

From Fig. \ref{fig2} it is clear that the cumulative average depends
on the initial time and that after a time stabilizes into a value that
it is difficult to change unless that there is a big swing in opinion
of the electorate. Thus, while this indicator captures the primary
election well, it is quite static and resilient to any change that
could happen long after the initial time of measurement. To investigate
this effect, we have recalculated the cumulative average by changing
the origin of measurement $t_0$ and plot in Fig. \ref{fig3} the
different cumulative predictions for different values of $t_0$. We see
that the results at the day of the election clusters well around the
results of the PASO primaries when we consider different $t_0$. This
result indicates that the cumulative indicator is stable in the
present Argentinian election.

Using this extended cumulative model, our best possible prediction for
the upcoming general election on October 27, 2019 is that the
formula Fern\'andez-Fern\'andez will win by 16\% difference over
Macri-Pichetto.

\section{Conclusion}

One of the fundamental tools of AI in social networks is that it
captures changes in people's opinions without any intervention and for
an extended time. These millions of users who constantly express
themselves on the internet and change or maintain their positions now
have a new ally: the AI that captures people's popular sentiment,
filters it from manipulators and bots and reduces it to its
essence. For this reason, no traditional survey will come to
understand these positions with such precision. That is the
fundamental difference between modern techniques and traditional
pollsters regarding the monitoring of the vote or the opinion of
society.

The results of this analysis show, on the one hand, that AI applied to
big-data can be used to understand the large movements of opinion that
arise globally. On the other hand, traditional surveys will be
replaced by new methods based on modern prediction technologies such
as AI. Who does not understand these new digital trends, will end in
political and / or economic ostracism. AI is a thermometer that
provides the key to predicting not only the elections but the great
trends that develop at the local and global levels. Today, AI allows
to synthesize the opinion of millions of people who would not be heard
otherwise. We must not ignore that people are tired of answering
surveys. AI can then deduce, predict, interpret and understand what
people want to express.

\clearpage

\clearpage

\centerline{Supplementary Materials}

\section{Instantaneous and Cumulative average}
\label{def1}

% User-level: FF(u) represents whether user $u$ supports FF, dating from
% $d-w+1$ to $d$. She/He posted $n_{F,t}$ tweets classified to FF and
% $n_{t,M}$ to MP on $t$. Note that if $D-w<0$, $t$ starts from 1.

% \begin{equation}
% FF(u)=\left\{
% \begin{aligned}
% 1 & , & \sum_{t=d-w+1}^{d}n_{F,t} > \sum_{t=d-w+1}^{d}n_{M,t} \\
% 0 & , & \sum_{t=d-w+1}^{d}n_{F,t} = \sum_{t=d-w+1}^{d}n_{M,t} \\
% -1 & , & \sum_{t=d-w+1}^{d}n_{F,t} < \sum_{t=d-w+1}^{d}n_{M,t} \\
% \end{aligned}
% \right.
% \end{equation}

% Country-level: $Nu_{d}(FF=1)$ users would vote for FF, $Nu_{d}(FF=1)$
% would vote for MP and $Nu_{d}(FF=0)$ users are undecided or vote for
% other candidates. We can obtain the percentage of voters in different
% camps from:

% \begin{equation}
%   FF_{d} = \frac{Nu_{d}(FF=1)}{Nu_{d}}
% \end{equation}

% \begin{equation}
%   MP_{d} = \frac{Nu_{d}(FF=-1)}{Nu_{d}}
% \end{equation}

% \begin{equation}
%   Others_{d} = \frac{Nu_{d}(FF=0)}{Nu_{d}}
% \end{equation}

% \section{Definition of UltraLoyals, and Loyal leaning voters}
% \label{def2}

For instantaneous prediction, we define three categories of users where M = Macri tweet, F = Fernandez tweet.
Users posted $n_{M,t}$ tweets classified as M 
and $n_{F,t}$ tweets classified as FF on $t$.
We compare the number of M and F dating from $T-w+1$ to $T$ ($w=14$ in this study).
Note that if $T-w<0$, $t$ starts from 1 (the first day of observation). 

\begin{itemize}

\item Classified as $MP$:

\begin{equation}
M_T(u)=\left\{
\begin{aligned}
1, & \,\,\,\,\, {\mbox if} \,\,\,\,\, \sum_{t=T-w+1}^{T}n_{M,t} > \sum_{t=T-w+1}^{T}n_{F,t} \\
0, & \,\,\,\,\, {\mbox otherwise} \\
\end{aligned}
\right.
  \end{equation}

\item Classified as $FF$:
\begin{equation}
F_T(u)=\left\{
\begin{aligned}
1, & \,\,\,\,\, {\mbox if} \,\,\,\,\, \sum_{t=T-w+1}^{T}n_{M,t} < \sum_{t=T-w+1}^{T}n_{F,t} \\
0, & \,\,\,\,\, {\mbox otherwise} \\ 
\end{aligned}
\right.
\end{equation}

\item Classified as $Undecided$:
\begin{equation}
 \,\,\,\,\,\,\,\,\,\,\, U_T(u)=\left\{
\begin{aligned}
1, & \,\,\,\,\, {\mbox if} \,\,\,\,\, \sum_{t=T-w+1}^{T}n_{M,t} = \sum_{t=T-w+1}^{T}n_{F,t} > 0\\
0, & \,\,\,\,\, {\mbox otherwise} \\ 
\end{aligned}
\right.
\end{equation}

\end{itemize}

We give the instantaneous prediction (percentages) on $T$-th day, according to the number of users that belong to different categories above.

\hspace*{\fill}

For cumulative prediction, we define four categories of users where M = Macri tweet, F = Fernandez tweet.

\begin{itemize}

\item Classified as $MP$:

\begin{equation}
M_T(u)=\left\{
\begin{aligned}
1, & \,\,\,\,\, {\mbox if} \,\,\,\,\, \sum_{t=T_0}^T n_{M,t} > \sum_{t=T_0}^T n_{F,t} \\
0, & \,\,\,\,\, {\mbox otherwise} \\
\end{aligned}
\right.
  \end{equation}

\item Classified as $FF$:
\begin{equation}
F_T(u)=\left\{
\begin{aligned}
1, & \,\,\,\,\, {\mbox if} \,\,\,\,\, \sum_{t=T_0}^T n_{M,t} < \sum_{t=T_0}^T n_{F,t} \\
0, & \,\,\,\,\, {\mbox otherwise} \\ 
\end{aligned}
\right.
\end{equation}

\item Classified as $Undecided$:
\begin{equation}
 \,\,\,\,\,\,\,\,\,\, U_T(u)=\left\{
\begin{aligned}
1, & \,\,\,\,\, {\mbox if} \,\,\,\,\, \sum_{t=T_0}^T n_{M,t} = \sum_{t=T_0}^T n_{F,t} > 0 \\
0, & \,\,\,\,\, {\mbox otherwise} \\ 
\end{aligned}
\right.
\end{equation}

\item $Unclassified$ users do not enter into the categories.
They talk about the candidates FF and MP or others candidates 
but they do not express their intention to vote for MP nor FF.

\end{itemize}

Besides, We define $third\ party$ category (also called $Others$) that includes both $Undecided$ and $Unclassified$ users.

We give the cumulative prediction (percentages) on $T$-th day, according to the number of users that belong to different categories above.


\clearpage


\begin{thebibliography}{\bibliographystyle}

  
 \expandafter\ifx\csname url\endcsname\relax
 \def\url#1{\texttt{#1}}\fi
\expandafter\ifx\csname urlprefix\endcsname\relax\def\urlprefix{URL }\fi
\expandafter\ifx\csname doiprefix\endcsname\relax\def\doiprefix{DOI }\fi
\providecommand{\bibinfo}[2]{#2}
\providecommand{\eprint}[2][]{\url{#2}}

% 1
\bibitem{surveys} R. Tourangeau, F. G. Conrad and M. P. Couper,
 M. P. {\it The Science of Web Surveys} (Oxford University Press, New
 York, 2013). 
 \url{https://www.aapor.org/AAPOR_Main/media/MainSiteFiles/Sampling-Methods-for-Political-Polling_1.pdf}

% 2
\bibitem{jacobs} Jacobs, J., and B. House (2016), 'Trump Says He
 Expected to Lose Election Because of Poll Results,' Bloomberg.com,
 December 13, 2016. Retrieved from
 \url{https://www.bloomberg.com/politics/articles/2016-12-14/trump-says-he-expected-to-lose-election-because-of-poll-results}

% 3
\bibitem{pew} Pew research, 27/02/2019. Retrieved from
\url{https://www.pewresearch.org/fact-tank/2019/02/27/response-rates-in-telephone-surveys-have-resumed-their-decline/}

% 4
\bibitem{aapor} C. Kennedy {\it et al.} Ad Hoc Committee on 2016
 Election Polling, {\it An Evaluation of 2016 Election Polls in the
  U.S.} American Association for Public Opinion Research, AAPOR
 report (2016). Retrieved from
 \url{https://www.aapor.org/Education-Resources/Reports/An-Evaluation-of-2016-Election-Polls-in-the-U-S.aspx}

% 5
\bibitem{erraron} Clar\'in, 27/09/2019. Retrieved from (in Spanish,
 translation available online from Google Translate):
 \url{https://www.clarin.com/politica/encuestadoras-fuego-erraron-paso-dicen-octubre_0_T72H9hdl.html}

% 6
\bibitem{clarin} Clar\'in, 15/08/2019. Retrieved from (in Spanish,
 translation available online from Google Translate):
\url{https://www.clarin.com/opinion/intrigas-casa-rosada-pases-factura-city-lunes-negro_0_jnggAIsh5.html} 
% \url{https://www.baenegocios.com/politica/La-encuesta-que-rompio-una-consultora-20190812-0053.html})

% 7
\bibitem{wsj} The Wall Street Journal, 05/09/2019. Retrieved from
\url{https://www.wsj.com/articles/hedge-fund-loses-1-billion-in-one-month-on-argentina-bet-11567696547}

% 8
\bibitem{bovet}
 A. Bovet, F. Morone, H. A. Makse, {\it Validation of Twitter
 opinion trends with national polling aggregates: Hillary Clinton vs
 Donald Trump}, {\bf Sci. Rep.} {\bf 8673} (2018).
  
% 9
\bibitem{fake}
  A. Bovet, H. A. Makse, {\it Influence of fake news in Twitter
 during the 2016 US presidential election}, {\bf Nature
 Comm. 10}, 7 (2019).

% 10
\bibitem{book}
 A. Bovet, S. Pei, F. Morone, H. A. Makse, {\it The Science of
 Influencers Using Mathematically Rigorous Theories - Understanding
 the Future of Society, Biology, Markets and Ecosystems} (Springer
 Nature, Switzerland, 2019) forthcoming.

% 11
\bibitem{jasny} B. Jasny and R. Stone, {\it Prediction and its
limits}, Science {\bf 355} (Feb 3, 2017) (cover feature).

% 12
\bibitem{nytimes}
\bibinfo{author}{{New York Times}}.
\newblock \bibinfo{title}{{New York Times National Polling Average}}
 (\bibinfo{year}{2016}).
\newblock
 \urlprefix\url{http://www.nytimes.com/interactive/2016/us/elections/polls.html}.
 \newblock \bibinfo{note}{[Online; accessed 14-Oct-2019]}.

% 13
\bibitem{wiki_election}
 Wikipedia contributors, 18/10/2019. Retrieved from
 \url{https://en.wikipedia.org/w/index.php?title=Opinion_polling_for_the_2019_Argentine_general_election&oldid=921809639}

% not used
% \bibitem{bohannon} J. Bohannon, {\it The Pulse of the People}.
%  Science {\bf 355}, 470-472 (2017).

% \bibitem{patent} 11/01/2018. US Patent App. 16/019075 (2018). Title:
% {\it Method to maximize message spreading in social networks and
%  find the most influential people in social media}. Inventors:
% Hern\'an Makse and Flaviano Morone.

% \bibitem{riesgo} Clar\'in, 04/19/2019. Retrieved from (in Spanish,
%  translation available online from Google Translate)
%  \url{https://www.clarin.com/economia/revuelo-encuesta-ayudo-subir-riesgo-pais_0_qsecQJtQc.amp.html}

% \bibitem{manipulation} Impulso Baires, 10/12/2019. Retrieved from (in Spanish,
%  translation available online from Google Translate):
 
%  \url{https://www.impulsobaires.com.ar/nota/275336/bomba-inversores-demandarian-a-elypsis-por-la-encuesta-del-viernes-y-los-k-cargarian-con-denuncia-penal-para-investigar-a-operadores}

% \bibitem{seido} Perfil, 10/21/2019. Retrieved from (in Spanish,
%  translation available online from Google Translate):

%  \url{https://www.perfil.com/noticias/bloomberg/bc-ex-elypsis-luciano-cohan-funda-nueva-consultora-en-argentina.phtml}

%  \bibitem{w1} Michael P. Battaglia, David Izrael, David C. Hoaglin,
%   and Martin R. Frankel. Tips and Tricks for Raking Survey Data
%   (a.k.a. Sample Balancing). American Association for Public
%   Opinion Research. JSM, 4740-4745 (2004).

%  \bibitem{w2} David Izrael, David C. Hoaglin, and Michael
%   P. Battaglia. A SAS Macro for Balancing a Weighted Sample.
%   Statistics and Data Analysis, Paper 258-25 (2000).
  
% \bibitem{land} New York Times, 10/04/2019. Retrieved from
%  \url{https://www.nytimes.com/2019/10/04/opinion/sunday/trump-arkansas.html?action=click&module=Opinion&pgtype=Homepage}

% \bibitem{kitsak} M. Kitsak, L. K. Gallos, S. Havlin, F. Liljeros,
%  L. Muchnik, H. Eugene Stanley, H. A. Makse, Identification of
%  influential spreaders in complex networks, Nat. Phys. {\bf 6}, 888 (2010).

% \bibitem{flaviano}
%  F. Morone, H. A. Makse. Influence maximization in complex networks through optimal percolation, Nature {\bf 524}, 65-68 (2015).

% \bibitem{Hopkins2010}
% \bibinfo{author}{Hopkins, D.~J.} \& \bibinfo{author}{King, G.}
% \newblock \bibinfo{journal}{\bibinfo{title}{{A method of automated
%  nonparametric content analysis for social science}}}.
% \newblock {\it{Am. J. Pol. Sci.}}
%  \textbf{\bibinfo{volume}{54}}, \bibinfo{pages}{229--247}
%  (\bibinfo{year}{2010}).
% \newblock \doiprefix 10.1111/j.1540-5907.2009.00428.x.

% \bibitem{Ceron2015}
% \bibinfo{author}{Ceron, A.}, \bibinfo{author}{Curini, L.} \&
%  \bibinfo{author}{Iacus, S.~M.}
% \newblock \bibinfo{journal}{\bibinfo{title}{{Using sentiment analysis to
%  monitor electoral campaigns: method matters--evidence from the united states
%  and Italy}}}.
% \newblock {\emph{Soc. Sci. Comput. Rev.}}
%  \textbf{\bibinfo{volume}{33}}, \bibinfo{pages}{3--20} (\bibinfo{year}{2015}).
%  \newblock \doiprefix 10.1177/0894439314521983.
 
% %\bibitem{Ceron2016}
% %\bibinfo{author}{Ceron, A.}, \bibinfo{author}{Curini, L.} \&
% % \bibinfo{author}{Iacus, S.~M.}
% %\newblock \bibinfo{journal}{\bibinfo{title}{{ISA: A fast, scalable and accurate
% % algorithm for sentiment analysis of social media content}}}.
% %\newblock {\emph{Inf. Sci. (Ny).}}
% % \textbf{\bibinfo{volume}{367-368}}, \bibinfo{pages}{105--124}
% % (\bibinfo{year}{2016}).
% % \newblock \doiprefix 10.1016/j.ins.2016.05.052.

% \bibitem{OConnor2010}
% \bibinfo{author}{O'Connor, B.}, \bibinfo{author}{Balasubramanyan, R.},
%  \bibinfo{author}{Routledge, B.~R.} \& \bibinfo{author}{Smith, N.~a.}
% \newblock \bibinfo{title}{{From tweets to polls: Linking text sentiment to
%  public opinion time series}}.
% \newblock \bibinfo{pages}{122--129} (\bibinfo{year}{2010}).
% \newblock \doiprefix citeulike-article-id:7044833.

% \bibitem{Marchetti-bowick2012}
% \bibinfo{author}{Marchetti-Bowick, M.} \& \bibinfo{author}{Chambers, N.}
% \newblock \bibinfo{title}{{Learning for microblogs with distant supervision:
%  political forecasting with Twitter}}.
% \newblock In \emph{\bibinfo{booktitle}{Proc. 13th Conf. Eur. Chapter
%  Association Comput. Linguist.}}, \bibinfo{pages}{603--612}
%  (\bibinfo{year}{2012}).

% \bibitem{Thapen2013}
% \bibinfo{author}{Thapen, N.~A.} \& \bibinfo{author}{Ghanem, M.~M.}
% \newblock \bibinfo{title}{{Towards passive political opinion polling using
%  twitter}}.
% \newblock In \emph{\bibinfo{booktitle}{CEUR Workshop Proc.}}, vol.
%  \bibinfo{volume}{1110}, \bibinfo{pages}{19--34} (\bibinfo{year}{2013}).

%  \bibitem{Jungherr2016}
% \bibinfo{author}{Jungherr, A.}, \bibinfo{author}{Schoen, H.},
%  \bibinfo{author}{Posegga, O.} \& \bibinfo{author}{Jurgens, P.}
% \newblock \bibinfo{journal}{\bibinfo{title}{{Digital trace data in the study of
%  public opinion: an indicator of attention toward politics rather than
%  political support}}}.
% \newblock {\emph{Soc. Sci. Comput. Rev.}}
%  \bibinfo{pages}{0894439316631043} (\bibinfo{year}{2016}).
 
% \bibitem{scikit-learn} \bibinfo{author}{Pedregosa, F.} \emph{et~al.}
%  \newblock \bibinfo{journal}{\bibinfo{title}{Scikit-learn: machine
%    learning in {P}ython}}. \newblock {\emph{J. Mach. Learn. Res.}}
%  \textbf{\bibinfo{volume}{12}}, \bibinfo{pages}{2825--2830}
%  (\bibinfo{year}{2011}).
 
% \bibitem{urban} Pew Internet, 04/24/2019. Retrieved from     
% \url{https://www.pewinternet.org/2019/04/24/sizing-up-twitter-users/}.

%  \url{https://ropercenter.cornell.edu/how-groups-voted-2016}
 
% \bibitem{fake2} N. Salamanos, M. J. Jensen, X. He, Y. Chen,
%  M. Sirivianos. On the Influence of Twitter Trolls during the 2016
%  US Presidential Election. \url{https://arxiv.org/abs/1910.00531v1}

\end{thebibliography}
\end{document}